# @C – augmented version of C programming language

Iosif Iulian Petrila[*]

Faculty of Automatic Control and Computer Engineering, Gheorghe Asachi Technical University of Iasi, Str. Dimitrie Mangeron, Nr. 27, 700050, Iasi, Romania

The augmented version of C programming language is presented. The language was completed with a series of low-level and high-level facilities to enlarge the language usage spectrum to various computing systems, operations, users. The ambiguities and inconsistencies have been resolved by managing problematic and undefined languages elements through an interpretation and management similar to that used in the case of other C syntax based languages. The proposed augmentative completeness elements, through @C approach, preserve the spirit of C language and its basic characteristics through compatibility with the standard version but also allow rejuvenation and bring C language to the present programming languages state of the art.

**Keywords:** @C language, C language, programming languages, programming instruments, system instruments, languages inconsistencies, languages completeness

**Introduction**

There is a master/ultimate rule in nature according to which all objects (microscopic or macroscopic) in the universe are transformed according to a minimum of action, which means with a minimum effort in relation to the complexity or resources required or used. The same rule we apply in different fields if we want to be efficient, in fact we implicitly applying Occam's razor rule used for the validation of any scientific approach: it is unnatural (useless/ineffective) to achieve with more what we can do with less. In computing, this rule was extremely strictly used especially in the early years of computing mainly from constraints related to resource limitations in terms of memory and computing power. Over the time, however, the rule began to be ignored, thus some IT tools becoming extremely ineffective, especially when we try to scale or adapt them to new technologies. Looking back in time, through the 1970s, when computer systems began to grow, the related tools (operating systems, but especially programming languages) also began to develop, but most of them in a non-optimal way, under different influences etc. However, among these, although initially neglected and even marginalized by the IT community because it came with a series of concepts and operations that did not find an immediate mathematical correspondent (such as the pointer with the related arithmetic operations, increment/decrement operations etc.), the closest to the principle of efficient management of the computing system resources was the C programming language [1]. Even though it was initially designed as a development language for the famous operating system – UNIX (alias of UNICS – UNiplexed/Multiplexed Information Computing System/Service) [2], the C language appeared and stood out as a tool which perform an optimal management of the computer system resources, with facilities far beyond of other languages of those times and which allowed the easy description of various operations very close to low level form but also with high level characteristics, even if those descriptions did not find an adequate mathematical expression at that time. The representativeness of the C language for computer systems is given by the fact that at that time it achieved the most relevant symbiosis between the two categories of elements that a programming language must integrate and generalize in its syntax: processor instructions and operating/administrative system methods. Later, with the abundance of computing resources, many non-optimized tools began to be used, including the C language begins to evolve, not always in an appropriate manner.
There is no doubt that the C language [1] is and probably will remain the most representative system programming language, being a benchmark to the direct language successors (C++, C#, D etc.) and also to languages based on its syntax (C like languages), representing in fact a class reference to almost all programming languages [3-7]. Over the time, many of its elements were included by both high-level (user-oriented) and low-level (machine-oriented) languages but not vice versa, as it should be for a middle level language to include at least the most representative elements from both levels. Consequently, being still the most important/representative computing tool, the C language (and related compilers) must be updated in accordance with today's technologies challenges with preservation of C spirit and flexibility. First of all, the language must facilitate to a greater extent the management of both low-level elements (because the variety of computing devices is increasing) and high-level (by including functional computing elements, asynchronous management routines such as closures, web technologies facilities etc.). Secondly, the language must become self-consistent in the sense of containing those elements necessary for a self-compilation (even as an elementary test of completeness) without using other languages or tools (for instance make file routines should be embedded in the language in preprocessing and postprocessing directives etc.), to allow correlations between compilation stages (preprocessing, processing/encoding, linking, postprocessing), to facilitate just-in-time compilation etc [8]. All of these can be achieved without turning it into a new language, but only from a more rigorous and consistent management of language elements, specific to a dialect, through augmentations and flexibilisations, with minimal additions so that the vast majority of codes written so far to be managed properly. A flexibilisation of the C language is also necessary from the architectural perspective in order to release the language from the CISC fingerprint (especially x86, historically explainable) and to facilitate the easy management of RISC/embedded systems as well as other emerging computing systems: quantum, neuromorphic etc [9-21].

---

[*] **E-mail address:** IosifIulianPetrila@gmail.com

**@C Characteristics**

A goal of the @C approach/dialect/vision is to give to the C language the natural flexibility, in "the spirit of C", as fine adjustments, with minimal changes or completions and with a natural correlation with the structure of computing systems in the idea of facilitating an easy approach (with clear rules) of the both extremely low and high levels of processing as it is normal for a middle level language to include at least the most representative elements from other levels, but at the same time to keep the compatibility with the latest usual versions of the C programming language. Also, those additions should include the current specifics of computing systems for which computing resources are no longer critical / limited at the developer level etc. Another target is to include in the language the directives currently used by compilers as external elements (often as paralanguages or different languages elements, which for historical reasons would be understandable), but inherently present in the process of generating the compilation product. From general user/programmer perspective, it is extra inconvenient/effort to learn other languages to perform certain extra operations that could be very simply implemented in the C language, this will be possible through @C augmentation. Last but not least, this approach can be considerate as 50 years celebration of C and also represent a tribute to the pioneers of the early stages of the computer systems tools development, especially to Dennis MacAlistair Ritchie and Kenneth Lane Thompson, because at the system level the current development tools are today essentially what they designed, in fact, the computing system development is still based on their achievements. These pioneers were the ones who provided a natural approach to computer system programming, with specific distinct implementative operations (as a natural operate with a numerical quantity which within in computing systems has a real physical correspondent on a finite number of bits stored at an address in memory which must be accessed and which is not an mathematical abstract concept) allowing the computing field to emancipate itself as an independent domain. However, over the time, once with the mathematization of IT concepts, it is important to use rules as clear as possible at the language level (this can only be achieved through the rigor of mathematics and not through the endless management of exceptions) to avoid inconsistencies, ambiguities, errors etc. Deviation from rules is understandable in situations where significantly and more efficient implementations are obtained (such as placing some variables at aligned addresses in some cases/architectures etc.), but if the efficiency is insignificant then the rules must prevail (such as evaluating the arguments of procedures in the order in which they appear regardless of calling convention type etc.). In the following the basic characteristics of @C will be highlighted, with some relevant arguments, solutions, examples etc.

The comments are usually substituted with space in conventional C language compilers, before or at the beginning of preprocessing stage. This substitution is not always appropriate because it allows the fragmentation of the tokenization process even when it not should be, but, for historical reasons, the syntactic-lexical characteristics are preserved, only that in the case of @C, the comments are processed similarly to any other token, they are only equivalent to spaces and not explicitly substituted, in this way, both types of comments are managed more appropriate and used similar to other languages. Multi-line or block comment starts with `/*` if these characters are not placed in a string or in a single-line comment and ends with `*/` no matter where these characters are placed, even in a string or in a single-line comment.

```
/*
  This is a block
  or multi-line
  comment!
*/
```

The single-line comment starts with `//` if these characters are not placed in a string or a multi-line comment and ends at end of line.

```
//This is a single-line comment!
```

Similar to other languages based on C syntax, single line comment continuation (which ends with line-continuation escape character \) is excluded in @C

```
//Single-line comment ended with \
printf("Line also commented?\n");
```

The basic punctuators used in @C are the same as in C language, except that they are accepted without ambiguous combinations or in some combinations for which rules based on priorities are needed, rules that are not immediately obvious to the simple user. For example, the expressions like `i+++j` will be considered ambiguous and treated with a notification/warning or an error in order to guide the user to be more rigorous and to write it as he intended as `i++ +j` or `i+ ++j`. This approach will give to the codes a more lisible description, in addition to the aesthetic aspect, being also in accordance with good practices which suggest successive operators separated by space etc.

Regarding tokens and identifiers content, along with string literals, the @C language is flexible because facilitate the programmer descriptions of some elements in any natural language. In this respect, it is useful to accept identifiers described in different languages (in some dialects/compilers this is restricted)

```
int Number, Număr, 数字, संख्या, Число;
```

because often for the common user the names are also semantic indications of some language elements in the absence of related comments (which can be accessible to any language) etc. The advantages of restrictions (such as a stricter control of errors etc.) must be balanced with the advantages of flexibility (such as an easier and more concise use of the language by the global programmers etc.).

Even if, for general cases, two types of control instructions are needed [8], in the initial and final stages of the compilation of C language it is enough to complete and make flexible the control instructions from the preprocessing stage and to be used even for output formation decisions, optimizations etc. In order to achieve this, it is essential that the preprocessing stage to be interspersed with the coding stage, this intercalation will not essentially affect the preprocessing of the previously macro descriptions. In other words, the specific elements of the preprocessing stage must be combined with the language elements, even if (for historical reasons) they were introduced and managed almost independently of the language itself. The C language was structured to be processed in one logical compilation step/pass through the explicit pre-declarative structure. Therefore, in @C approach is establishing two-way correlations between the preprocessing stage and the compilation/coding stage. For this purpose, on the one hand the control instructions are completed with other types of instructions (such as repetitive or cycling instructions as `#while`) [8], besides the standard ones `#ifdef / #ifndef / #if / #else / #enidf` and on the other hand, their conditions are also allow the verification of some characteristics specific to the coding process such as if a parameter/variable has been `declared`, `coded`, `used` etc, in addition to the macro specific verification such as `defined` which will keep its preprocessing meaning (for compatibility), its counterpart for pure/codable language (non macro) identifiers being `coded`, obviously with the similar meaning. In this way, due to the interaction and alternations of the preprocessing operations with coding ones in @C, the preprocessing control instructions can be used in the coding stage, even in code optimization operations. For instance, if a `declared` procedure it may not be `used` (called / invoked / referred / assigned), then one may use a conditional inclusion/exclusion of its forward defining/coding area:

```
#if used P
  void P(){/*..*/}
#endif
```

which is useful to eliminate the code/body of unused/uncalled procedure etc. These interactive (macro and coding) facilities are not incompatible with previous implementations of preprocessing operations that were considered (and implemented in the usual compilers) distinct from coding operations, but only make them more flexible, through their use even in the coding stage, the compilation processes and decisions becomes explicitly accessible to language descriptions without using specifications in other extra languages to perform certain compiling operations etc.

Numeric literals are accepted within @C in various forms/bases for expressions compatibility with other languages. That including also the direct descriptions in base 2, neglected over the time by the C standard, although extremely useful to have direct access to the bits of a numerical value/parameter, especially from a low level perspective etc. Also, in order ensure compatibility with other C dialects and for readability of large numbers descriptions, underscore character `_` will be accepted for numeric literals fragmentation and grouping as: `0b1_0000_0000` for `0b100000000`, `1_000_000` for `1000000`, `0xFF_FF` for `0xFFFF` etc.

One of the limits of the C language in the operating with binary parameters and operations was the binary limitation of zero-terminated strings. These limitations are removed in @C through consideration of the declaration of an array without the number of elements explicitly specified as being to a variable length array managed as a dynamic array

```
int A[];
```

compatible with the usual static array, but for which the number of elements can be determined through `length` internal function. In this way, @C facilitates easy management of binary-save strings (which can contain `\0`)

```
char S[] = "Binary-safe\0 @C String!";
```

which can be displayed with

```
printf("%.*s", length(S), S);
```

but also through some standard versions viewed as a static array

```
printf("%.*s", sizeof(S) -1, S);
```

with some limitations related to access to the definition/allocation of `S` for a proper use of `sizeof` in the previous example, limitations that do not exist if `S` string is passed to another variable and the content is managed also as a variable length array (dynamic array) to which `length` can be applied

```
char V[] = S; printf("%.*s", length(V), V);
```

In the case of usual C (except for some versions), the declaration of an uninitialized array variable without the number of elements is accepted in some particular cases and these corresponding to static array declarations, which in @C are used in their explicit form. For example, as last element in a `struct` in order to map as an array the elements from the post structure address, only that in the case of @C this facility is accepted only in the explicit and correct form

```
struct S
{
  //..
  int L[0];
};
```

where `L` component is the correct and explicit definition for post `struct` data mapped with `L` field with 0 space allocate on `struct` for both @C and C. In this way must be declared and used a flexible array member of a `struct`, while member `A` from the following structure

```
struct S
{
  //..
  int A[];
};
```

will be managed as a dynamic array in @C, different from the C standard case for which the last two descriptions are equivalent, the last one being the inappropriate definition of a `struct` flexible array member.

There are a number of inconsistencies / ambiguities in how common C compilers evaluate expressions, ambiguities that are often inherited from early C compiler implementations and from the fact that the expression concept has evolved over time, inconsistencies which are an important source of errors, side effects, undefined/unpredictable behaviors/results etc. These originate primarily from the compiler's arbitrary unsequenced implementation of statements and expressions, the problem that can be solved by assuming sequencing and rule of parsing from left to right etc. For example, the current reader can first try to estimate what will display the following code and then can test it on different compilers

```
int I = 0; printf("%d, %d\n", ++I + ++I, I++ + I++);
```

most likely obtaining different values than estimated values and also other than the values expected (estimated) by common programmers. The correct and consistent implementation should print the values `3, 5`, identical to the values obtained by similar codes described in other languages that inherit the C syntax (such as Java, JavaScript etc.). The present @C version of C languages solve even these ambiguities by more rigorously defining the pre or post increment and decrement operations, the evaluation order of the arguments of a procedure etc.

The sequence point ambiguity on the usual C expressions can be highlighted from the analysis of the following example

```
int I = 0; int E = ++I + I++ + ++I + I++;
```

for which common C compilers give different results `E = 6`, `8`, `9` etc. In order to be portable, these types of expressions should be evaluated identically in any language. From this perspective, in @C, pre/post increments/decrements operations are managed similarly to those in the other languages for which the related operations are performed relative to the local context of their occurrence within the expression and not relative to the statement or expression block which are not always well defined etc. In this way, in @C case, as in other languages based on C syntax, the correct and predictable/unambiguous result for the previous example expression is `E = 8`.

An important issue, less visible to common users, is the order of arguments evaluations in procedures, illustrated by the following simple example

```
int I = 0; printf("%d, %d\n", I++, I++);
```

for which the most C compilers displays `1, 0` which is not `0, 1` as common programmers expected to be. The problem is inherited by compilers based on an erroneous implementation according to which the evaluation of arguments from right to left, as they are also saved on stack in the case of standard C, is faster and this speed is more important than the rule of logical parsing/processing from left/first to right/last of procedure arguments. In fact, at that time, right-to-left processing of arguments was a convenient choice implementation and not a speed issue, the speed was invoked to argue this unjustified deviation from the rule. Clear rules and restrictions on expression evaluation order are essential. Rules are more important (reduce errors) than a little bit faster but unpredictable code generation (with a few irrelevant percentage) related to some difficulties or inabilities in some particular systems implementations. Over time, the implementation was maintained for reasons of compatibility with previous implementations and later considered as an unspecified behavior seen as a freedom left to compilers to achieve faster implementations etc. In order to solve this problem, the default @C calling and evaluations of arguments is a new one named `ccall`, in which the arguments are evaluated from left to right and saved onto stack from right to left (as `cdecl`), unlike the common C case which use `cdecl` as the default procedure calling convention in which the arguments are also evaluated from right to left. These aspects can be easily highlighted through the following procedure

```
void P(int A, int B)
{
  printf("%c-%c\n", A + 'A', B + 'A');
}
```

called with

```
int I = 0; P(I++, I++);
```

which will print `A-B` with proposed @C `ccall` as `P(0, 1)` and `B-A` with the usual `cdecl` convention as `P(1, 0)`. The evaluation from the last argument to the first (`cdecl`) can often introduces errors that are difficult to be identify/track by the common user, because the arguments are not processed naturally in the order in which they appear in the enumeration (call) list as the common programmer expected to be and also as it is in other languages.

Similar to higher-level languages or other modern system languages, the @C language allows the definition of local procedures, in both (pure) nested (**N**) and closure (**C**) forms, as in the following example

```
ResultType Parent(int A)
{
  int V = 0;

  volatile void N(){/*..*/}

  void C(){/*..*/}

  //..

  //ExternCVar = C;

  //return C;
}
```

in which the **volatile** directive indicates that **N** is managed as nested procedure (safe to call only until the parent procedure return) and local procedures without **volatile** directive as **C** will be closure procedure (safely exportable and callable at any time with the contextual preservation of parent used environment this meaning preservation of used parent/ascendants arguments and local parent/ascendants variables). Along with the easy management at the parent block level of some local procedures, this facility will allow to use the language for higher level operations, asynchronous exchange of information through a save contextual callbacks etc. Along with some high-level procedural features, in @C, a series of related high-level data and operations are also accepted in order to facilitate some concise description, non-critical operations etc.

On the low level side, the **asm** directive allows in @C the use in an easier way of machine-type descriptions in addition to the assembly ones. Some preprocessing directives are correlated with the coding directives also under the low-level operations aspect. Other directives used in different versions or proposed (such as **#embed**) are integrated into easy resource descriptions related to some system formats etc. Also, in order to facilitate connections between low-level blocks, the definition and use of labels and related **goto** instruction will be made more flexible, allowing easy implementations of various low level optimizations such as recursive tail calling procedures etc.

The purpose of a programming language is to mediate the description of some operations from the human to the machine level through the fundamental translator within computing systems which is the compiler [8]. In @C language, the way in which the related compiler manages a series of operations is used explicitly at the language level and not just as some directives of the compiler as in the case of the conventional C language. However, as today many of the systems are interconnected, the @C language facilitates the description of some related operations not only at the library level but also at the internal and compiler level. For example, the main script server functions must not be exclusively specific to script languages but also to programming languages, in this sense, the @C language and related compiler also work as a (script) server by generalizing the source concept to accept some inputs not only as files but also as ports, addresses etc. The @C present augmented version of C language by including the most representative low and high level elements bring the language to the current programming languages state of the art, in accordance with the original language structure, in the spirit of C.

**Conclusions**

The proposed augmented version of the C language allows the flexibilisation and optimization of the language in a general way with facilities that keep the update compatible with the standard language version.

The language include the low level elements that allow a more adequate management of hardware components and embedded systems but also include the representative high level languages elements that allow an easier coding from the user's perspective.

The ambiguities and inconsistencies have been resolved to make the language compatible with languages that use the same type of syntax by eliminating unpredictive or unexpected behaviors or results, following to a greater extent the use of rules and the avoidance of exceptions, even if some exceptions to the rules could generate a little bit faster codes in some particular systems.

A series of elements have been introduced that will facilitate the direct compilation of codes, allowing the implementation of single tool compiler systems able of managing resources and generating applications installation archives, without other paralanguages, tools, dependencies, other auxiliary languages etc.